\documentstyle[prd,aps,epsfig,preprint]{revtex}
\tighten
\begin{document}
\draft
\title{The Toroid Moment of Majorana Neutrino}
\author{Vladimir~M.~Dubovik\protect\thanks{E-mail:
        dubovik@thsun1.jinr.dubna.su}}
\address{Bogoliubov Laboratory of Theoretical Physics, \\
         Joint Institute for Nuclear Research, Dubna 141980, Russia}
\author{Valentin~E.~Kuznetsov\protect\thanks{E-mail:
        valya@nu.jinr.dubna.su or valya@axnd02.cern.ch}}
\address{Laboratory of Nuclear Problems, \\
         Joint Institute for Nuclear Research, Dubna 141980, Russia}
\date{\today}
\maketitle
\thispagestyle{empty}
\begin{abstract}
If neutrino is the Majorana particle it can possess only one 
electromagnetic characteristic, the toroid dipole moment (anapole) 
in the static limit and nothing else. We have calculated the diagonal
toroid moment (form factor) of the Majorana neutrino by the 
dispersion method in the one-loop approximation of the Standard Model
and found it to be different from zero in the case of massive as well
as massless neutrinos. All external particles are on the mass shells
and there are no problems
with the physical interpretation of the final result. 
Some manifestations of the toroid interactions of Majorana 
neutrinos, induced by their toroid moments, are also remarked.
\end{abstract}
\pacs{12.15.Lk, {\bf 13.10.+q}, {\bf 13.15.+g}, 13.40.Gp}
\narrowtext
\begin{center}{\bf I. INTRODUCTION} \end{center}
The electromagnetic properties of Dirac and 
Majorana neutrinos are the subject of great interest at 
present~\cite{Reviews}. The difference between massive Dirac and
massive Majorana neutrinos is clearly exhibited by their
electromagnetic properties. As early as 1939, Pauli remarked that
the Majorana neutrinos have neither a magnetic dipole 
moment nor an electric dipole moment in vacuum~\cite{Pauli}.
Nevertheless the electromagnetic properties of Dirac and Majorana
neutrinos can manifest themselves via the anapole moment also 
\cite{Kayser1982}.

The anapole moment of $\frac{1}{2}$-spin Dirac particle 
was introduced by 
Zel'dovich~\cite{Zel1957} for a T-invariant interaction which does
not conserve P-parity and C-parity individually. Subsequently, a more
convenient characteristic was pointed out to describe of this kind of
interaction, the toroid dipole moment (TDM) \cite{Dubovik1974}. As
was shown, the TDM is a general case of the anapole, it coincides
with an anapole on the mass-shell of the particle under consideration
and has a simple classical analog. Similar to an electric dipole and
a magnetic dipole moments the TDM is the first term of third 
multipole family, the toroid moments. This 
type of static multipole moments does not produce any external 
electromagnetic fields in vacuum but generates a free-field 
(gauge-invariant) transverse-longitudinal 
potential~\cite{Dubovik_K} which
is responsible for topological effects such as the 
Aharonov -- Bohm effect. As was pointed out in~\cite{Radescu1985}
the Majorana particles of any value of the spin are characterized 
by toroid moments and nothing else.

A calculation of the vacuum anapole moment (TDM) of Dirac particle
was started in \cite{Aydin1972}. Then, a number of articles about
the problems of renormalizability, gauge non-invariance and,
consequently, observability of the neutrino anapole moment and 
neutrino charge radius (NCR) was published. 
(See~\cite{Aydin1972,Czyz1988,NCR,Dubovik1983} and 
references therein.)
However, as was pointed out in \cite{Gongora}, these quantities
are finite and well-defined in the Standard Model (SM) as being
the axial-vector (TDM) and the vector (NCR) contact interactions
with an external electromagnetic field, respectively.

In this paper we calculate the vacuum diagonal TDM (form factor) of 
the Majorana neutrino in the framework of the SM. In Sec. 2, we start 
by clarifying the reason why the Majorana neutrinos can possess only 
one electromagnetic characteristic, the TDM. We also discuss
the distinctions and similarities of the anapole and TDM of 
$\frac{1}{2}$-spin
particle and correctly define the TDM of the Majorana neutrino. 
Using some examples in Sec. 3, we illustrate the toroid interactions 
of the Majorana neutrinos. Further in Sec. 4, using the dispersion 
method, we calculate the TDM of the Majorana neutrino in the one-loop
approximation of the SM. All external particles are on the mass
shells and there are no problems with the physical interpretation of
the final result.We summarized our results in Sec. 5.
\begin{center}
{\bf II. THE TOROID DIPOLE MOMENT OF THE MAJORANA NEUTRINO}
\end{center}
In the case of Majorana neutrinos, the amplitude of the interaction
with an external electromagnetic field ${\cal A}^\mu$, 
${\cal M}\propto eJ^{\rm EM}_\mu(q){\cal A}^\mu(q)$ is defined by the
particle and antiparticle contributions
 \begin{eqnarray}
 J^{\rm EM}_\mu(q)&=&\Bigl[
 \overline{u}_f({\bf p}')\Gamma_\mu(q) u_i({\bf p})+
 \overline{v}_i({\bf p})\Gamma_\mu(q) v_f({\bf p}')\Bigr]
 \nonumber \\
                  &\equiv&
 \overline{u}_f({\bf p}')\left[\Gamma_\mu(q)
-\left(C^{-1}\Gamma_\mu(q)C\right)^T\right]u_i({\bf p}),
 \label{II.1}\end{eqnarray}
where $e$ is the charge of an electron, $q_\mu=p_{\mu}'-p_\mu$ is the
transferred 4-momentum, $u(p)$ and $v(p)$ are bispinors and $C$ is 
the charge conjugation matrix (we will use the chiral 
representation of the gamma matrices with $C=i\gamma_0\gamma_2$ 
and the normalization $\overline{u}({\bf p})u({\bf p})=1$); here
$\Gamma_\mu(q)$ is the electromagnetic vertex, which is characterized
by a set of electromagnetic form
factors~\cite{Kayser1982,Dubovik1974}. 
One popular way of defining the Lorentz structure of 
$\Gamma_\mu(q)$ is as follows:
 \begin{eqnarray}
  \Gamma_\mu(q) = F(q^2)\gamma_\mu 
               &+&  M(q^2)\sigma_{\mu\nu}q^\nu
                   +E(q^2)\sigma_{\mu\nu}q^\nu\gamma_5
                \nonumber \\
               &+&A(q^2)[q^2\gamma_\mu-\widehat{q}q_\mu]\gamma_5,
 \label{el.vtx}\end{eqnarray}
where $F$, $M$, $E$ and $A$ are called the normal magnetic, anomalous
magnetic, electric, and anapole dipole form factors, respectively.
These form factors are physically observable quantities as 
$q^2\rightarrow 0$ and 
their combinations define the well-known magnetic 
$(\mu)$, electric $(d)$ 
and anapole $(a)$ dipole moments. In the non-relativistic limit, 
the energy of interaction with an external electromagnetic field has 
the following form:
 \[{\cal H}_{\rm int}\propto
                       -\mu\left(\mbox{\boldmath $\sigma$}
                                 \cdot{\bf B}\right)
                         -d\left(\mbox{\boldmath $\sigma$}
                                 \cdot{\bf E}\right)
                         -a\left(\mbox{\boldmath $\sigma$}
                                 \cdot{\rm curl}\,{\bf B}\right), \]
where {\bf B} and {\bf E} are the strengths of the magnetic and
electric fields. In the case of Majorana neutrinos, imposing the
restriction of the CPT-invariance and using the C-, P-, T-properties
of $\Gamma_\mu(q)$ and ${\cal H}_{\rm int}$, which we have combined
in Table \ref{CPT}, we see that the magnetic and electric dipole
moments are absent in the static limit when the masses of the initial
$m_i$ and final $m_f$ neutrino eigenstates are equal to each other.
This means that Majorana neutrinos possess only one electromagnetic
characteristic, the anapole moment \cite{Kayser1982,Zel1957}.
But as was pointed out in Ref. \cite{Dubovik1974}, the anapole moment
does not have a simple classical analog -- $A(q^2)$ does not
correspond to certain multipole distribution and, therefore, a more
convenient characteristic, the TDM, was proposed to describe the
T-invariant interaction with non-conservation of P and C symmetries.
For clarity, we rewrite the axial part of the
electromagnetic vertex (\ref{el.vtx}) in the multipole 
parametrization~\cite{footnote1}
 \begin{eqnarray}
 \Gamma_\mu(q) &\propto& \Bigl\{ 
                        i\varepsilon_{\mu\nu\lambda\sigma}
                         P^\nu q^\lambda\gamma^\sigma\gamma_5 T(q^2)
                        +\sigma_{\mu\nu}q^\nu D(\Delta m^2)
                         \nonumber \\
               & - &     \frac{q^2P_\mu-(q\cdot P)q_\mu}
                         {q^2-\Delta m^2}
                         \left[D(q^2)-D(\Delta m^2)\right]
                         \Bigr\}\gamma_5,
 \label{el.vtx2}\end{eqnarray}
where $\varepsilon_{\mu\nu\lambda\sigma}$ is the Levi-Civitta unit 
antisymmetric tensor, $P_\nu=p_{\nu}'+p_{\nu}$, $\Delta m=m_i-m_f$ 
and $D(\Delta m^2)$, $D(q^2)$ and $T(q^2)$ are the charge dipole 
moment, charge dipole, and toroid dipole form factors, respectively. 
In this parametrization, there is a one-to-one correspondence
in the definition of multipole moments by their form factors: the
electric dipole moment by $D(\Delta m^2)$, the TDM by 
$T(\Delta m^2)$, etc. That is not a case in (\ref{el.vtx}), where,
for instance, the electric dipole moment
$$   d\propto ie\Bigl[E(\Delta m^2)-\Delta m A(\Delta m^2)\Bigr]   $$
is defined in terms of the electric and anapole dipole form factors. 
Using the following identities:
 \begin{eqnarray} 
    \overline{u}_f({\bf p}')\Bigl\{ q^2\sigma_{\mu\nu}q^\nu
 &+&\Delta m\left(q^2\gamma_\mu-\widehat{q}q_\mu\right) 
    \nonumber \\ 
 &+&\left[q^2P_\mu-\left(q\cdot P\right)q_\mu\right]
                  \Bigr\}\gamma_5 u_i({\bf p}) = 0,
    \nonumber \\ 
    \overline{u}_f({\bf p}')\Bigl\{ \Delta m\sigma_{\mu\nu}q^\nu
 &+&\left(q^2\gamma_\mu-\widehat{q}q_\mu\right) 
    \nonumber \\ 
 &-&i\varepsilon_{\mu\nu\lambda\sigma}P^\nu q^\lambda
                  \gamma^\sigma\gamma_5
                  \Bigr\}\gamma_5 u_i({\bf p}) = 0, 
  \label{II.3}\end{eqnarray}
we obtain the connection between the anapole and TDM
 \begin{equation}
 A(q^2) = T(q^2) + \frac{m_i^2-m_f^2}{q^2-\Delta m^2}
                   \left[D(q^2)-D(\Delta m^2)\right].
 \label{AM-TDM}\end{equation}
As can be seen, they coincide only on the mass-shell of the particle
under consideration. The definition of the anapole by two independent
form factors leads to confusion in the classical limit and gives no
way for an analytical continuation of this form factor on the
mass-shell \cite{Dubovik1974}. Hence, the TDM is a more convenient 
electromagnetic characteristic of the particle than the anapole.

Using the standard definitions of dipole moments \cite{Semikoz1989},
we define the TDM of the Majorana neutrino as a pseudovector
${\bf T}$ that is directed along the spin of the particle, the only
vector characteristic in its own rest frame,
 \[ T_\mu  = eT(0)\overline{u}(0)\gamma_\mu\gamma_5 u(0),  \quad
    {\bf T}= eT(0)\varphi^{\dag}\mbox{\boldmath $\sigma$}\varphi,  \]
where $\varphi$ is the Pauli spinor. 
In the coordinate representation, {\bf T} is the total moment of the
following density distribution:
 \[ g({\bf r})=\frac{1}{10}[{\bf r}({\bf Jr})-2r^2{\bf J}],        \]
where ${\bf J}$ is the current density that produces the 
magnetic field. The interaction of TDM with 
an external electromagnetic field has the following form
\cite{PerZel}:
 \begin{eqnarray} 
    {\cal H}_{\rm int} & = &
    eT(q^2)\overline{N}(x)\left[q^2\gamma_\mu{\cal A}^\mu(x)
        -\gamma_\mu q^\mu q_\nu{\cal A}^\nu(x)\right] \gamma_5 N(x)
                       \nonumber \\
                       & = & 
    eT(q^2)\overline{N}(x)\gamma_\mu\gamma_5 N(x)
 \left[ \frac{\partial^2{\cal A}^\mu(x)}
             {\partial x^\nu\partial x_\nu}
       -\frac{\partial^2{\cal A}^\nu(x)}
             {\partial x^\nu\partial x_\mu}
 \right]
                       \nonumber \\
                       & = &
    eT(q^2)\overline{N}(x)\gamma_\mu\gamma_5 N(x)
        \frac{\partial F^{\mu\nu}(x)}{\partial x^\nu},
 \label{TI1}\end{eqnarray}
where  $N(x)$ is the Majorana neutrino field, which has a 
usual plane wave expansion, and satisfies the following condition
$$     N^c(x)=CN(x)C^{-1}=C\overline{N}^T(x)=N(x).              $$
$F^{\mu\nu}(x)$ is the tensor of the electromagnetic field, which,
in turn, produces an external current 
$$    \partial_\nu F^{\mu\nu}(x)=-J^\mu_{\rm EM}(x).            $$
It is easy to see that in the nonrelativistic limit, when the
particle is in its own rest system of reference,
 \[ q^2\rightarrow 0, \quad
    \overline{N} \gamma_0\gamma_5 N \rightarrow 0, \quad
    \overline{N} \mbox{\boldmath $\gamma$}\gamma_5 N \rightarrow 
    \varphi^{\dag}\mbox{\boldmath $\sigma$}\varphi,             \]
the corresponding interaction energy is
 \begin{equation} 
 {\cal H}_{\rm int}=-{\bf T}\cdot{\bf J}
                   =-eT(0)\varphi^{\dag}\mbox{\boldmath $\sigma$}
                          \varphi\left({\rm curl}\,{\bf B}
                                  -\dot{{\bf E}}\right).
 \label{TI2}\end{equation}
It has the moment of force
$$  {\bf M}={\bf T}\;[\mbox{\boldmath $\sigma$}\times{\bf J}],  $$
and represents a T-invariant toroid (anapole) interaction of the 
particle which does not conserve P-parity and C-parity individually,
and defines the axial-vector contact interaction with an external
electromagnetic field \cite{Zel1957,Gongora}.
\begin{center}{\bf III. REMARKS ON SOME PROPERTIES OF THE
               TOROID INTERACTIONS OF MAJORANA NEUTRINOS}
\end{center}
The TDM has many different applications, both in classical 
electrodynamics and in quantum 
theories~\cite{Dubovik1974,Dubovik_K,Dubovik1983}.
The simplest model of TDM (anapole) was given by 
Zel'dovich~\cite{Zel1957} 
as a conventional solenoid folded into a torus and having only a 
poloidal current. For such a stationary solenoid, having neither an
azimuthal (toroidal) component of the current nor electric fields
around the torus, there is only a nonzero azimuthal magnetic field
inside the torus. Therefore, fields outside the permanent toroid
dipole are zero in vacuum, as well. However, as was pointed out by
Ginzburg and Tsytovich \cite{Ginzburg1985}, when the toroid dipole
moves in a medium,
the latter might be regarded as permitting the dipole itself and 
the fields outside the dipole appear to produce, for instance,  
Vavilov-Cherenkov radiation. Also, the toroid dipole is responsible 
for the transition radiation when it goes through the interface 
between two media whose indices of refraction are $n_1$ and $n_2$ 
($n_1 \gg n_2$). However, we should stress here that the result of 
\cite{Ginzburg1985} was derived in the framework of classical
electrodynamics and it might change in quantum theory with the 
consideration of the Vavilov-Cherenkov and transition radiations of 
neutrinos induced by their toroid moments moving in the medium.

The toroid interactions of Dirac or Majorana neutrinos manifest 
themselves in collisions of the neutrinos with charged particles
where the TDM conserves the helicity of the neutrino and gives an
extra contribution, as a part of the radiative corrections, to the
total cross section of 
the scattering of neutrinos by electrons, quarks and nuclei. 
In this regard, the toroid moment is similar to NCR. Both conserve 
the helicity in coherent neutrino scattering, but have different
natures. They define the axial-vector (TDM) and the vector (NCR)
contact interactions with an external electromagnetic field.
Such interactions are the subject of interest in low-energy
scattering processes and give one way to probe the NCR and TDM. 
For further implications and references in the question of the
observability of NCR, see Refs. \cite{NCR,Allen}.

The toroid interactions of neutrinos may have a very interesting 
consequences in different media. As was pointed out 
in \cite{Semikoz1989}, the electromagnetic properties of Dirac and
Majorana neutrinos in the medium are similar to the vacuum case. 
For instance, the Majorana neutrinos do not have the induced electric
and magnetic dipole moments but have an induced anapole (toroid)
dipole moment. However, it has a nonzero value in an anisotropic
medium such as ferromagnetic material and absent in an isotropic
medium. Nevertheless the vacuum TDM may play a very important role
itself, in particular, for neutrino oscillations. For instance, let
us consider the evolution equation for three neutrino flavors
$\vec{\nu} = (\nu_e,\nu_\mu,\nu_\tau)^T$ in the presence 
of toroid interactions
 \[ i\frac{d\vec{\nu}}{d\tau}=
    K\left[\frac{1}{2E}{\rm diag}
     \left(m_1^2,m_2^2,m_3^2\right)+{\cal T}(\tau)
     \right]K^{\dag}\vec{\nu},                                    \]
where $K$ is the mixing matrix connecting the flavor basis, 
$\nu_\ell$ $(\ell=e,\mu,\tau)$, and the
mass basis, $N_i$ $(i=1,...,k)$, of Majorana neutrinos as 
$\nu_\ell=\sum_i K_{\ell i} N_i$,
and has the $3(k+1)$ mixing angles and $3(k+1)$ CP-violating phases 
(for details see \cite{Schechter1980}.) The matrix ${\cal T}$ is, 
in general, a $3\times3$ matrix whose elements
 \begin{equation}
  {\cal T}_{if}(\tau) \propto T_{if}\,\,
   \mbox{\boldmath $\sigma$}\cdot{\rm curl}\,{\bf B}(\tau),
 \label{eq1}\end{equation}
are functions of time $\tau$ different from zero in the presence of 
the inhomogeneous vortex magnetic field which, in a concrete 
experimental situation, may be realized according to Maxwell's
equations as a displacement current or the current of the particle
colliding with the considered neutrino at the space point where the
interaction (\ref{eq1}) is determined.

In this sense, this problem is an analog of the well-known 
Wolfenstein equation for the propagation of neutrinos through a 
medium \cite{Wolfenstein78}, but resonance conversion of neutrinos
can occur even in vacuum, due to the toroid interactions of neutrinos 
(if ${\cal T}_{ii}\neq{\cal T}_{jj}$)~\cite{footnote2}.
The off-diagonal matrix elements ${\cal T}_{if}$, 
induced by the transition toroid moments, are nontrivial factors in 
the Wolfenstein equation and had no previous analogs in the SM 
(beyond the scope of SM this role was played by the so-called flavor 
changing neutral currents). Since the Hamiltonian of evolution of the 
three neutrino flavors contains at least one time-dependent varying 
external parameter 
$\mbox{\boldmath $\sigma$}\cdot{\rm curl}\,{\bf B}(\tau)$, 
we should take into account the topological phases in the evolution
operator \cite{Berry}, which may be very important
for neutrino oscillations \cite{Naumov1}. If, an addition, a neutrino
beam intersects some fluctuations of density and element compositions
in the background of matter, a new phenomenon, geometric resonance,
in neutrino oscillations occurs \cite{Naumov2}. The role of the two 
time-dependent parameters of the Hamiltonian (varying independently
from one another), which are need for the geometric resonance, can be
played by the external electromagnetic field (in the medium, it can
be the electron current and/or intrinsic sources) and the medium
itself ($n_{\nu_\ell}\neq1$). For example, if the
${\rm curl}\,{\bf B}(\tau)$ and particle number density $\rho(\tau)$
vary cyclically when a neutrino beam propagates through the medium,
i.e., ${\rm curl}\,{\bf B}(\tau)={\rm curl}\,{\bf B}(0)$ and 
$\rho(\tau)=\rho(0)$
for some time $\tau$, they form a closed contour on the plane
($\mbox{\boldmath $\sigma$}\cdot{\rm curl}\,{\bf B}$, $\rho$), and
for some neutrino momentum the geometric resonance takes place
(for details, see \cite{Naumov2}.)

The transitions in a system of Majorana neutrinos with anapole and
transition magnetic moments propagating in matter with a twisting
non-potential magnetic field have recently been investigated within 
the asymmetric left-right model by Boyarkin and 
Rein \cite{BoyarkinRein}. It has been shown that the resonance
conversion of neutrinos appears not only in response to the influence
of matter, but also due to the availability of electromagnetic
moments.

These effects can manifest themselves in numerous astrophysical and
cosmological situations. Among them are neutrino propagation through
the solar interior and young supernova envelopes, neutrino radiation
of accreting neutron stars and black holes, where inhomogeneous 
vortex magnetic fields can have large values, etc.,
and the toroid interactions of neutrinos should be taken 
into account in each of them. 
But, the conclusions about the magnitude of these effects 
have been the subject of separate investigations, which is beyond
the scope of our present work.

Since there is great interest in neutrino properties at the moment,
we present here the calculation of the diagonal TDM, $T_i(0)$, of
the mass eigenstate, $N_i$, of the Majorana neutrino in the framework
of the SM.
\begin{center}{\bf IV. ONE-LOOP RESULT}\end{center}
The TDM of the Majorana neutrino can be defined in the one-loop 
approximation of the SM of electroweak interactions from the Feynman
graphs shown in Figs. \ref{Fig1}, \ref{Fig2}. As one can see from 
(\ref{el.vtx}) and (\ref{II.3}), the transition TDM is equal to the 
diagonal one plus the part proportional to the neutrino mass
difference. Therefore, to calculate the diagonal TDM and to estimate
the transition one of the Majorana neutrino, only the anapole 
parametrization is sufficient.

To illustrate, we shall give some details of our calculations for the
two graphs with $\ell\ell W$ states, see Fig. \ref{Fig3}. The
amplitudes and contributions to the imaginary part of the toroid form
factor for the other diagrams are given in Appendix B. It is easy to
verify that contributions of the particle and antiparticle currents
are equal to each other, and we will consider one of them, and thus
multiply the amplitude by a factor of 2. Using the Feynman rules and
notation summarized in Appendix A, we can write the amplitude in the
following form:
 \begin{eqnarray}
 {\cal M}&=&2\int\frac{d^4k_1}{(2\pi)^4}\frac{d^4k_2}{(2\pi)^4}
             (2\pi)^4\delta^4\Bigl[p_1-p_2-(k_1-k_2)\Bigr]
 \nonumber \\
         &\times &
             \overline{u}(p_1)\,\left[
             i\Gamma^{(\ell N)}_\lambda\right]\,
             i\Delta_F(k_1) 
             (-ie\gamma_\mu)\,i\Delta_F(k_2)\,
 \nonumber \\
         &\times &
             \left[i\overline{\Gamma}^{(\ell N)}_\nu\right]\,u(p_2)
             \,\left[i\Delta_W^{\lambda\nu}(p_1-k_1)\right]
             {\cal A}^\mu(k_1-k_2).
 \label{III.3}\end{eqnarray}
For convenience in our calculations, we pass to the $t$-channel where
the momenta of particles transform as
\[p_1\rightarrow p_-,\quad p_2\rightarrow -p_+,                     \]
\[k_1\rightarrow k_-\equiv k_1,\quad k_2\rightarrow -k_+\equiv -k_2,\]
\[t=q^2=(k_1+k_2)^2=(p_-+p_+)^2,                                    \]
and using the transformation
 \begin{equation}
 \frac{1}{k^2-m^2} \rightarrow (-2\pi i)\delta\left(k^2-m^2\right)
 \Theta(k_0),
 \label{III.4}\end{equation}
which is valid when we take into account the unitary condition for the
S-matrix \cite{LifPit,AxBer}, we can write the imaginary part of the 
amplitude as
 \begin{eqnarray} {\rm Im}\,\,{\cal M}&=&-e\int d\tau
                  \frac{{\cal A}^\mu(q)} {(p_--k_1)^2-m_W^2}
 \nonumber \\                     &\times &
                  \overline{u}(p_-)\Gamma^{(\ell N)}_\lambda
                  \left(\widehat{k}_1+m_\ell\right)\gamma_\mu 
 \nonumber \\                     &\times &
                  \left(\widehat{k}_2-m_\ell\right)
                  \overline{\Gamma}^{(\ell N)}_\nu 
                  g^{\lambda\nu}v(p_+).
 \label{III.5}\end{eqnarray}
Here, we have denoted the two-body phase-space factor as
 \begin{eqnarray*} d\tau&=&\frac{1}{(2\pi)^2}d^4k_1 d^4k_2
                           \delta^4(p_-+p_+-k_1-k_2) \\
                        &\times &
                           \delta\left(k^2_1-m^2_\ell\right)
                           \Theta(k_{10})
                           \delta\left(k^2_2-m^2_\ell\right)
                           \Theta(k_{20}).
 \end{eqnarray*}
Now, keeping only the terms with $\gamma_\mu\gamma_5$, and using the 
identity
 \[ {\rm Im}\,\,{\cal M}=e\overline{u}(p_-)\,{\rm Im}\,\Bigl[
    T_i(t)(t\gamma_\mu-\widehat{q}q_\mu)\gamma_5\Bigr]v(p_+)
    {\cal A}^\mu(q),                                                \]
and the gauge $q_\mu{\cal A}^\mu=0$, we perform the
two-particle phase space integration, following 
Refs.~\cite{Aydin1972}, and obtain the contribution to the imaginary
part of the diagonal toroid form factor for the $\ell\ell W$ diagrams:
 \begin{eqnarray} 
  {\rm Im}\,\,tT_i(t)  &   =   & \frac{
  \left|A^{(\ell N)}_L\right|^2-\left|A^{(\ell N)}_R\right|^2
                                    }{16\pi} 
  \Bigl(L_\ell-I_\ell-J_\ell\Bigr),
 \label{III.6}\end{eqnarray}
where
 \begin{eqnarray}
  L_k & = & \frac{1}{\lambda}\ln\left|\frac{1+b_k}{1-b_k}\right|, 
  \quad
  J_k   =   \frac{\sqrt{a_k}}{\lambda^2}\Bigl(2-\lambda b_kL_k\Bigr),
  \nonumber \\
  I_k & = & a_k\left[\frac{b_k}{\lambda}+
                     \frac{1}{2}(1-b_k^2)L_k\right],
  \quad \lambda = \sqrt{1-\frac{4m_i^2}{t}},
  \nonumber \\
  a_k & = & \left(1-\frac{4m_k^2}{t}\right), \quad
  b_k   =   \frac{a_k+2(m_W^2+m_\ell^2-m_i^2)/t}
                 {\lambda\sqrt{a_k}}.
  \nonumber \\
 \label{notations}\end{eqnarray}
The real part of the toroid form factor can be derived by using the 
dispersion relation with one subtraction \cite{LifPit},
 \begin{equation}
 tT_i(t)-\phi=\frac{t}{\pi}\int_{4m_\ell^2}^\infty
   \frac{{\rm Im}\,\,t'T_i(t')}{t'(t'-t-i0)}dt',
 \label{disp}\end{equation}
where $\phi$ is some constant, which we put zero in agreement with
(\ref{el.vtx}). Since 
$T_i(t)\rightarrow {\rm Const}\neq\infty$ as $t\rightarrow0$, we 
calculate the real part of the toroid form factor for $t\leq0$, where 
$T_i(t)={\rm Re}\,\,T_i(t)$.
Introducing the new variable 
$x=\frac{t'}{2m_W^2}$ and putting $m_i=0$, for
simplicity, we obtain
 \begin{equation}
 T_i(t) = \frac{
        \left|A^{(\ell N)}_L\right|^2-\left|A^{(\ell N)}_R\right|^2
               }{32\pi^2m_W^2}
        \int_{2\beta}^{\infty}\frac{F(x,\beta)dx}{x(x+\alpha)},
 \label{III.7}\end{equation}
where $\alpha=-\frac{t}{2m_W^2} > 0$ and the integrand reads
 \begin{eqnarray*}
 F&   =   &\left(\frac{\beta-1}{x}-3\right)\sqrt{1-\frac{2\beta}{x}}
 \nonumber \\
  &   +   &\left[2\left(1+\frac{1}{2x}\right)^2
               -\frac{\beta}{x^2}\left(1+x-\frac{\beta}{2}\right)
           \right] 
 \nonumber \\
       &\times &
                \ln\left[\frac{1+x-\beta+\sqrt{x(x-2\beta)}}
                              {1+x-\beta-\sqrt{x(x-2\beta)}}\right],
 \end{eqnarray*}
with $\beta=\frac{m_\ell^2}{m_W^2}$. Finally, using the
definition of the matrices $A^{(x)}_{L,R}$ and $B^{(x)}_{L,R}$,
see eq. (\ref{A3}), and performing elementary integrations for 
these two graphs and others (making appropriate expansions in 
$\frac{m_\ell^2}{m_W^2}$ and denoting 
$f=e,\mu,\tau, u, d, s, c, b, t$),
we obtain for $|t|=0$:
 \begin{eqnarray}
 T_i(0)             &=& \frac{\sqrt{2}G_F}{12\pi^2}
            \Biggl[
            C^i_{WWZ}+C^i_{W\phi Z}+C^i_{\phi\phi Z}+C^i_{ffZ}
 \nonumber \\       & & + \,
            \sum_{\ell=e,\mu,\tau}\Bigl(
            C^i_{\ell\ell W}+C^i_{\ell\ell\phi}+C^i_{WW\ell}
           +C^i_{W\phi\ell}+C^i_{\phi W\ell}
           +C^i_{\phi\phi\ell}\Bigr)
            \Biggr],
 \nonumber \\
 C^i_{\ell\ell W}   &=& |K_{\ell i}|^2\Biggl[
            \frac{11}{6}
           -\ln4\beta
           +\beta
            \left(\frac{7}{4}+\frac{\ln2}{2}
           -\frac{9}{8}\ln3\right)+{\cal O}(\beta^2)\Biggr], 
 \nonumber \\
 C^i_{\ell\ell\phi} &=& |K_{\ell i}|^2\beta
            \left[\frac{1}{6}-\frac{1}{2}
            \ln4\beta
           +{\cal O}(\beta)\right],
 \nonumber \\
 C^i_{WW\ell}       &=& |K_{\ell i}|^2
            \left[-\frac{5}{6}
           +\frac{7}{6}\beta
           +{\cal O}(\beta^2)\right],
 \nonumber \\
 C^i_{W\phi\ell}    &=& C^i_{\phi W\ell} = |K_{\ell i}|^2
            \left[-\frac{1}{8}\beta
           +{\cal O}(\beta^2)\right],
 \nonumber \\
 C^i_{\phi\phi\ell} &=& |K_{\ell i}|^2
            \left[-\frac{1}{12}\beta
           +{\cal O}(\beta^2)\right],
 \nonumber \\
 C^i_{ffZ}          &=&-\,\,\Omega_{ii}\sum_{f\neq t}(g_L^f+g_R^f)
            \left[
            \frac{8}{3}-c_f-\frac{3}{2}
            \sqrt{c_f}\left(1-\frac{c_f}{3}\right)
            \ln\left|\frac{1+\sqrt{c_f}}{1-\sqrt{c_f}}\right|
            \right]
 \nonumber \\
                    & &-\,\,\Omega_{ii}(g_L^t+g_R^t)
            \left[
            \frac{8}{3}+c_t-3
            \sqrt{c_t}\left(1+\frac{c_t}{3}\right)
            \arctan\frac{1}{\sqrt{c_t}}
            \right],
 \nonumber \\
 C^i_{WWZ}          &=&-\,\,\Omega_{ii}
            \left[
            \frac{77}{6}+2c_W-\frac{\sqrt{c_W}}{2}(27+4c_W)
            \arctan\frac{1}{\sqrt{c_W}}
            \right],
 \nonumber \\
 C^i_{W\phi Z}      &=&-\,\,\frac{3}{4}\Omega_{ii}\sin^2\theta_W
            \left[
            \frac{2}{3}+c_W-\sqrt{c_W}(1+c_W)
            \arctan\frac{1}{\sqrt{c_W}}
            \right],
 \nonumber \\
 C^i_{\phi\phi Z}   &=&-\,\,\frac{1}{2}\Omega_{ii}(1-2\sin^2\theta_W)
            \left[
            \frac{1}{3}-c_W\left(1-\sqrt{c_W}
            \arctan\frac{1}{\sqrt{c_W}}\right)
            \right],
 \label{III.8}\end{eqnarray}
where 
 \begin{eqnarray*}
    c_k      & = & \left|1-\frac{4m_k^2}{m_Z^2}\right|,
                   \quad k=f,W,                                   \\
    g_L^\ell & = & -\frac{1}{2}+\sin^2\theta_W,\quad
    g_R^\ell   =\sin^2\theta_W,                                   \\
    g_L^U    & = & \frac{1}{2}-\frac{2}{3}\sin^2\theta_W,\quad
    g_R^U      =-\frac{2}{3}\sin^2\theta_W,\quad U=u,c,t,         \\
    g_L^D    & = & -\frac{1}{2}+\frac{1}{3}\sin^2\theta_W,\quad
    g_R^D      =\frac{1}{3}\sin^2\theta_W, \quad D=d,s,b,
 \end{eqnarray*}
and $K_{\ell i}$, $\Omega_{ii}$ are elements of the mixing matrices 
$K$ and $\Omega$ (see Appendix A.) We have also considered the
$ff\phi$, $ff\phi^0$, $ccZ^0$, $WW\phi^0$, $\phi\phi\phi^0$ and
$cc\phi^0$ graphs, since they appear in the 't Hooft-Feynman gauge,
but their amplitudes do not contain terms with the 
$\gamma_\mu\gamma_5$ structure and, therefore, they do not contribute
to the TDM of the Majorana neutrino.

Summing over all contributions, we get
 \[
 T_i(0)  \approx  \frac{\sqrt{2}G_F}{12\pi^2}\!\!
                  \left[\sum_{\ell = e, \mu, \tau}|K_{\ell i}|^2
                  \left(1+\ln\frac{m^2_W}{4m^2_\ell}\right)
                  -P_i\right],
 \]
where 
$P_i=-\Bigl(C^i_{WWZ}+C^i_{W\phi Z}+C^i_{\phi\phi Z}+C^i_{ffZ}\Bigr)$ 
is the polarization-type contribution to the TDM. As one can see from 
(\ref{III.8}), the $C^i_{\ell\ell W}$, $C^i_{WW\ell}$ and $P_i$ 
contributions are the leading terms that define the TDM. Since the
quark mass values have various values in different hadronic models,
the $C^i_{ffZ}$ contribution contains large uncertainties. Taking
into account the masses of leptons, $W,Z$-bosons and 
limits on the ``current-quark masses'' \cite{PDG}, we 
obtain~\cite{footnote3}
 \begin{eqnarray}
 T_i(0) &\approx& 1.277\times 10^{-33}\Bigl(|K_{ei}|^2
                  +0.547|K_{\mu i}|^2
 \nonumber \\ 
        &       & \quad\quad\,\,\,
                  +\,\,0.307|K_{\tau i}|^2
                  -0.425\times10^{-2}P_i\Bigr)
                  \quad ({\rm cm}^2),
 \nonumber \\ 
 P_i    &  \in  & \Bigl[8.585\;{\rm to}\;10.870\Bigr]
                  \,\,\Omega_{ii}.
 \label{III.9}\end{eqnarray}
Our result shows some very interesting peculiarities of TDM. In 
particular, as one can see from (\ref{III.8}) and (\ref{III.9}):
({\it i}) the value of the diagonal TDM depends on the
masses of leptons, quarks, $W,Z$ bosons, and matrix elements of the
mixing matrices $K$ and $\Omega$; ({\it ii}) the diagonal TDM of the
massive as well as massless neutrino has a finite value and depends
only very slightly on the mass of neutrino, 
${\cal O}\left(\frac{m_i^2}{m_W^2}\right)$; 
({\it iii}) the strong dependence of the TDM from the fermion masses
shows it is a good tool to test the SM, as well as the new
generation of fermions and bosons.
\begin{center}{\bf V. SUMMARY}\end{center}
While a Dirac neutrino has three dipole moments, a Majorana neutrino
possesses only one electromagnetic characteristic, in the static 
limit, the toroid (anapole) dipole moment. Nevertheless, the Majorana
neutrinos can have nonzero transition magnetic, electric and toroid
dipole moments. We have calculated the diagonal toroid dipole moment
of the mass eigenstate of the Majorana neutrino in the one-loop
approximation of the SM. It determines by the matrix elements of the
mixing matrices $K$ and $\Omega$, leptons, quarks and $W,Z$ bosons
masses and has the absolute value of the order of
$10^{-33}-10^{-34}$ ${\rm cm}^2$. 
This value is very sensitive to uncertainties of quark masses 
which define the up and down limits of TDM of the mass eigenstate
of the Majorana neutrino, see eq. (\ref{III.9}). We also found
that TDM has a finite value in the case of massive as well as massless
neutrinos. If there is no mixing in the lepton sector, $K=\Omega=1$,
we can define the singular electromagnetic characteristics, the
toroid moments, of the three weakly interacting massless neutrinos as:
 \begin{eqnarray} 
 T_{\nu_e}(0)    &\approx&
 \Bigl[+\,6.873\;\;{\rm to}\;+8.112\Bigr] \times 10^{-34}
 \quad ({\rm cm}^2),
                     \nonumber \\
 T_{\nu_\mu}(0)  &\approx&
 \Bigl[+\,1.090\;\;{\rm to}\;+2.329\Bigr] \times 10^{-34}
 \quad ({\rm cm}^2),
                     \nonumber \\
 T_{\nu_\tau}(0) &\approx&
 \Bigl[-\,1.971\;\;{\rm to}\;-0.732\Bigr] \times 10^{-34}
 \quad ({\rm cm}^2).
 \label{III.10}\end{eqnarray}

Toroid interactions are a part of the radiative corrections to the 
scattering of neutrinos on electrons, quarks and nuclei both in
vacuum and in medium. Therefore it gives a way to extract the 
experimental information about the magnitude of neutrino toroid 
moments in the low-energy scattering processes. 
Since the toroid interactions manifest themselves in the presence of 
an external electromagnetic field, they should be taken into account 
in various astrophysical and cosmological situations. As an example,
in this paper we are touched the problem of neutrino oscillations. We
are noticed that the toroid interactions can mediate the resonance
conversion of neutrinos even in vacuum in the presence of 
time-dependent external electromagnetic field. Also, they may have a
very interesting consequences on neutrino oscillations in the medium.
In fact, when the neutrino flux travels in the medium with a
time-dependent non-potential vortex magnetic field at some
conditions the geometrical resonance in neutrino 
oscillations occurs. However the conclusions about the magnitude
of these effects requires a separate investigation.

\acknowledgments

We are grateful to V. A. Naumov for numerous discussions 
and useful comments in reading of first versions of this paper.

\newpage
\begin{center}{\bf APPENDIX A}\end{center}

Here we give a short list of the Feynman rules used in our
calculations. Weak interactions of Majorana neutrinos $N$ and
charged leptons $\ell$ with gauge bosons $W^{\pm}$, $Z^0$,
non-physical scalars $\phi^{\pm}$, $\phi$ and Higgs particles
$\phi^0$ may be described 
by five Lagrangians
\cite{Gluza1992}:
 \begin{eqnarray*}
   {\cal L}_{\rm int}^{NW^{\pm}}    &=&
    \overline{N}\Gamma^{(\ell N)}_\mu\ell W^{+\mu}
   +\overline{\ell}\,\overline{\Gamma}^{(\ell N)}_\mu NW^{-\mu},
 \nonumber \\
   {\cal L}_{\rm int}^{NZ^0}        &=&
    \overline{N}\Gamma^{(N)}_\mu N Z^\mu
   +\overline{\ell}\Gamma^{(\ell)}_\mu\ell Z^\mu,
 \nonumber \\
   {\cal L}_{\rm int}^{N\phi^{\pm}} &=&
   \overline{N}\Gamma^{(\ell N)}\ell\phi^+
   +\overline{\ell}\,\overline{\Gamma}^{(\ell N)}N\phi^-,
 \nonumber \\
   {\cal L}_{\rm int}^{N\phi}       &=&
    \overline{N}\Gamma^{(N\phi)}N\phi
   +\overline{\ell}\Gamma^{(\ell\phi)}\ell\phi,
 \nonumber \\
   {\cal L}_{\rm int}^{N\phi^0}     &=&
    \overline{N}\Gamma^{(N)}N\phi^0
   +\overline{\ell}\Gamma^{(\ell)}\ell\phi^0.
 \end{eqnarray*}
The relevant Feynman rules are:
\begin{itemize}
\item $i\Gamma^{(\ell N)}_\mu$ for outgoing $W^-$ or incoming $W^+$,
\item $i\overline{\Gamma}^{(\ell N)}_\mu$ for outgoing $W^+$ or 
      incoming $W^-$,
\item $i\Gamma^{(\ell N)}$ for outgoing $\phi^-$ or incoming 
      $\phi^+$,
\item $i\overline{\Gamma}^{(\ell N)}$ for outgoing $\phi^+$ or
      incoming $\phi^-$,
\item $i\Gamma^{(N)}_\mu+i\Gamma^{(NC)}_\mu$ for $Z^0$,
\item $i\Gamma^{(N)}+i\Gamma^{(NC)}$ for $\phi^0$,
\item $i\Gamma^{(N\phi)}+i\Gamma^{(N\phi C)}$ for $\phi$.
\end{itemize}
Here 
\[ \Gamma^{(NC)}_\mu\equiv C\left[\Gamma^{(N)}_\mu\right]^TC^{-1},
   \quad
   \Gamma^{(NC)}    \equiv C\left[\Gamma^{(N)}\right]^TC^{-1},      \]
and similarly, for $\Gamma^{(N\phi C)}$.

As was pointed out in Ref. \cite{Gluza1992}, we should use the
following rule for the Dirac-Majorana transition in a Feynman graph
in real calculations: for an incoming (outgoing) Dirac particle, the
outgoing (incoming) Majorana neutrino must be treated as a particle,
and vice versa, for antiparticles.

Introducing the notation $P_{L,R}=\frac{1}{2}(1\mp\gamma_5)$, we can
write the general forms for all vertices
 \begin{eqnarray}
 \Gamma^{(x)}_\mu &=& \gamma_\mu
 \left[P_LA^{(x)}_L+P_RA^{(x)}_R\right],
  \quad x=\ell,N,\ell N\;, \nonumber \\
 \overline{\Gamma}^{(\ell N)}_\mu
                  &\equiv&
 \gamma_0\left[\Gamma^{(\ell N)}_\mu\right]^{\dag}\gamma_0=
 \gamma_\mu\left[P_LA^{(\ell N)*}_L+P_RA^{(\ell N)*}_R\right],
                           \nonumber \\
 \label{A1}\end{eqnarray}
and
 \begin{eqnarray}
 \Gamma^{(x)}     &=& P_LB^{(x)}_L+P_RB^{(x)}_R,\quad
  x=N,N\phi,\ell N, \nonumber \\
 \overline{\Gamma}^{(\ell N)}
                  &\equiv&
 \gamma_0\left[\Gamma^{(\ell N)}\right]^{\dag}\gamma_0=
 P_RB^{(\ell N)*}_L+P_LB^{(\ell N)*}_R.
 \label{A2}\end{eqnarray}
The other Feynman rules used in our calculations are well known and
are taken from \cite{LiCheng}. Equations (\ref{A1}--\ref{A2}) 
have a general form and must be specified for a given gauge group. 
Below, we present these matrices in the SM. 

We will use the following definitions of charged and neutral
currents:
 \[ J^-_\mu = \frac{1}{2}\overline{\ell}\gamma_\mu 
              (1-\gamma_5)K N, \quad
    J^0_\mu = \frac{1}{4}\overline{N}\gamma_\mu 
              (1-\gamma_5) \Omega N,                                \]
where $K$, in general, is a rectangular matrix, an analog of the 
Kobayashi-Maskawa matrix in the quark sector, such that $KK^{\dag}=1$
and $\Omega=K^{\dag}K\neq1$ \cite{Schechter1980}.
In the SM with 3 neutrino flavors from SU(2) doublets and 
$k$-singlets, the matrices $K$ and $\Omega$ have $3\times(3+k)$- and
$(3+k)\times(3+k)$-dimensions, respectively.
In this manner, we define the required matrices $A^{(x)}_{L,R}$ and
$B^{(x)}_{L,R}$ as:
 \begin{eqnarray}
 A^{(\ell N)}_L &=& \frac{g}{\sqrt{2}}K,\quad A^{(\ell N)}_R=0,
 \nonumber \\
 A^{(f)}_L      &=& \frac{gg^f_L}{\cos\theta_W}, \quad
                    A^{(f)}_R=\frac{gg^f_R}{\cos\theta_W},
 \nonumber \\
 A^{(N)}_L      &=& \frac{g\Omega}{2\cos\theta_W},\quad A^{(N)}_R=0,
 \nonumber \\
 B^{(\ell N)}_L &=&-\frac{gm_i K}{\sqrt{2}m_W}, \quad
                    B^{(\ell N)}_R=\frac{gm_\ell K}{\sqrt{2}m_W},
 \label{A3}\end{eqnarray}
with 
$$       G_F=\frac{\sqrt{2}g^2}{8m_W^2},                         $$
and use the propagators in the 't Hooft-Feynman gauge
 \begin{eqnarray*}
 i\Delta_F(k)              &=& \frac{i}{\widehat{k}-m_\ell
                             +i\epsilon},
 \\
 i\Delta^{\mu\nu}_{W,Z}(k) &=& \frac{-ig^{\mu\nu}}{k^2-M_{W,Z}^2
                              +i\epsilon},
 \\
 i\Delta_{\phi^\pm}(k)     &=& \frac{i}{k^2-m_W^2+i\epsilon},
 \\
 i\Delta_{\phi}(k)         &=& \frac{i}{k^2-m_Z^2+i\epsilon},
 \end{eqnarray*}

\newpage
\begin{center}{\bf APPENDIX B}\end{center}

In this appendix, we present the full list of amplitudes, in the 
$t$-channel, and the contributions to the imaginary parts of the
diagonal toroid form factor for triangular- and polarization-type
diagrams:

\begin{flushleft}{\bf $\ell\ell\phi$ triangular diagrams} 
\end{flushleft}
 \begin{eqnarray*}
 {\cal M}&=&2\int d\varrho\,\,
 \overline{u}(p_-)\,\left[i\Gamma^{(\ell N)}\right]\,
 i\Delta_F(k_1)\,(-ie\gamma_\mu)
 \nonumber \\ &&\times\,
 i\Delta_F(-k_2)\,
   \left[i\overline{\Gamma}^{(\ell N)}\right]\,v(p_+)
 \,\left[i\Delta_{\phi^+}(p_--k_1)\right]{\cal A}^\mu(q),
 \nonumber \\
 {\rm Im}\,\,tT_i(t)&=& \frac{1}{32\pi}
 \left(\left|B^{(\ell N)}_L\right|^2
-\left|B^{(\ell N)}_R\right|^2\right)
 \Bigl(I_\ell-J_\ell-L_\ell\Bigr),
 \end{eqnarray*}
where 
 \[ d\varrho = \frac{d^4k_1}{(2\pi)^4}\frac{d^4k_2}{(2\pi)^4}
            (2\pi)^4\delta^4(p_- + p_+ - k_1 - k_2).               \]

\begin{flushleft}{\bf $WW\ell$ triangular diagrams} \end{flushleft}
 \begin{eqnarray*}
 {\cal M}&=&2\int d\varrho\,\,
 \overline{u}(p_-)\left[i\Gamma^{(\ell N)}_\lambda\right]
 i\Delta_W^{\lambda\alpha}(k_1)
 \Bigl[-ieV_{\alpha\mu\beta}(q,-k_1,-k_2)\Bigr]
 \nonumber \\ &&\times\,
 i\Delta_W^{\beta\nu}(-k_2)
 \left[i\overline{\Gamma}^{(\ell N)}_\nu\right]v(p_+)
 \left[i\Delta_F(p_--k_1)\right]{\cal A}^\mu(q),
 \nonumber \\
 {\rm Im}\,\,tT_i(t)&=&
 \frac{1}{16\pi}
 \left(\left|A^{(\ell N)}_L\right|^2
-\left|A^{(\ell N)}_R\right|^2\right)
 \left( I_W-J_W+\frac{1}{2}a_WL_W-\frac{3}{2}L_W\right).
 \end{eqnarray*}
Here $V_{\alpha\mu\beta}(r_1,r_2,r_3)$ is the usual vertex function:
 \[ V_{\alpha\mu\beta}(r_1,r_2,r_3)=(r_1-r_2)_\beta\,g_{\mu\alpha}
    +(r_2-r_3)_\mu\,g_{\alpha\beta}+(r_3-r_1)_\alpha\,g_{\beta\mu},
    \quad r_1+r_2+r_3=0.
 \]

\begin{flushleft}{\bf $W\phi\ell$ triangular diagrams}
\end{flushleft}
 \begin{eqnarray*}
 {\cal M}&=&2\int d\varrho\,\,
 \overline{u}(p_-)\,\left[i\Gamma^{(\ell N)}_\lambda\right]\,
 i\Delta_W^{\lambda\alpha}(k_1)\,(iem_Wg_{\alpha\mu})
 \nonumber \\ &&\times\,
 i\Delta_{\phi^+}(-k_2)
 \left[i\overline{\Gamma}^{(\ell N)}\right]\,v(p_+)
 \,\left[i\Delta_F(p_--k_1)\right]{\cal A}^\mu(q),
 \nonumber \\
 {\rm Im}\,\,tT_i(t) &=&\frac{m_W}{16\pi t}
 \Biggl[ 
 \left( A^{(\ell N)}_RB^{(\ell N)\ast}_L
       -A^{(\ell N)}_LB^{(\ell N)\ast}_R
 \right)L_Wm_\ell
 \nonumber \\ &&+\,
 \left( A^{(\ell N)}_LB^{(\ell N)\ast}_L
       -A^{(\ell N)}_RB^{(\ell N)\ast}_R
 \right) (L_W+J_W)m_i
 \Biggr],
 \end{eqnarray*}

\begin{flushleft}{\bf $\phi W\ell$ triangular diagrams}
\end{flushleft}
 \begin{eqnarray*}
 {\cal M}&=&2\int d\varrho\,\,
 \overline{u}(p_-)\,\left[i\Gamma^{(\ell N)}\right]\,
 i\Delta_{\phi^+}(k_1) (iem_Wg_{\mu\beta})
 \nonumber \\ &&\times\,
 i\Delta_W^{\beta\nu}(-k_2)
 \left[i\overline{\Gamma}^{(\ell N)}_\nu\right]\,v(p_+)
 \,\left[i\Delta_F(p_--k_1)\right]{\cal A}^\mu(q), 
 \nonumber \\
 {\rm Im}\,\,tT_i(t) &=&\frac{m_W}{16\pi t}
 \Biggl[ 
 \left( B^{(\ell N)}_LA^{(\ell N)\ast}_R
       -B^{(\ell N)}_RA^{(\ell N)\ast}_L
 \right)L_Wm_\ell
 \nonumber \\ &&+\,
 \left( B^{(\ell N)}_LA^{(\ell N)\ast}_L
       -B^{(\ell N)}_RA^{(\ell N)\ast}_R
 \right) (L_W+J_W)m_i
 \Biggr],
 \end{eqnarray*}

\begin{flushleft}{\bf $\phi\phi\ell$ triangular diagrams}
\end{flushleft}
 \begin{eqnarray*}
 {\cal M}&=&2\int d\varrho\,\,
 \overline{u}(p_-)\,\left[i\Gamma^{(\ell N)}\right]\,
 i\Delta_F(p_--k_1)[-ie(k_2-k_1)_\mu]
 \nonumber \\ &&\times\,
 \left[i\overline{\Gamma}^{(\ell N)}\right]\,v(p_+)
 i\Delta_{\phi^+}(k_1)\,i\Delta_{\phi^+}(-k_2){\cal A}^\mu(q),
 \nonumber \\
 {\rm Im}\,\,tT_i(t) &   =   &
 \frac{I_W}{32\pi}
 \left(\left|B^{(\ell N)}_L\right|^2
      -\left|B^{(\ell N)}_R\right|^2\right).
 \end{eqnarray*}

\begin{flushleft}{\bf $ffZ$ polarization diagram} \end{flushleft}
 \begin{eqnarray*}
 {\cal M}&=&-\int d\varrho\,\,
 \overline{u}(p_-)\,
 \left[i(\Gamma^{(N)}_\nu+\Gamma^{(NC)}_\nu)\right]\,
 i\Delta_Z^{\nu\rho}(q) 
 \left[i\Gamma_\rho^{(f)}\right]
 \nonumber \\ &&\times\,
 i\Delta_F(-k_2)
 \Bigl(-ieQ_f\gamma_\mu\Bigr)
 i\Delta_F(k_1)v(p_+){\cal A}^\mu(q),
 \nonumber \\
 {\rm Im}\,\,tT_i(t) &=& \frac{1}{16\pi}
                         \left[A_R^{(N)}-A_L^{(N)}\right]
                         \left[A_R^{(f)}+A_L^{(f)}\right]
                         \sqrt{a_f}
                         \left(1-\frac{a_f}{3}\right)
                         \left(1-\frac{m_Z^2}{t}\right)^{-1}.
 \end{eqnarray*}

\begin{flushleft}{\bf $WWZ$ polarization diagram} \end{flushleft}
 \begin{eqnarray*}
 {\cal M}&=&\int d\varrho\,\,
 \overline{u}(p_-)\,
 \left[i(\Gamma^{(N)}_\nu+\Gamma^{(NC)}_\nu)\right]\,
 i\Delta_Z^{\nu\rho}(q)
 \Bigl[-ig\cos\theta_WV_{\rho\alpha\beta}(-q,k_2,k_1)\Bigr]
 \,i\Delta_W^{\alpha\alpha'}(k_1)
 \nonumber \\
 &&\times\,
 i\Delta_W^{\beta\beta'}(k_2)
 \Bigl[-ieV_{\mu\alpha'\beta'}(q,-k_1,-k_2)\Bigr]\,
 v(p_+){\cal A}^\mu(q),
 \nonumber \\
 {\rm Im}\,\,tT_i(t)&=& \frac{g\cos\theta_W}{96\pi}
 \left[A_R^{(N)}-A_L^{(N)}\right]\frac{(23t+16m_W^2)
 \sqrt{a_W}}{t-m_Z^2}.
 \end{eqnarray*}

\begin{flushleft}{\bf $W\phi Z$ polarization diagram}
\end{flushleft}
 \begin{eqnarray*}
 {\cal M}&=&\int d\varrho\,\,
 \overline{u}(p_-)\,
 \left[i(\Gamma^{(N)}_\nu+\Gamma^{(NC)}_\nu)\right]\,
 i\Delta_Z^{\nu\rho}(q) 
 \nonumber \\
 &&\times
 \left(-igm_Z\sin^2\theta_W g_{\rho\alpha}\right)
 \,i\Delta_W^{\alpha\alpha'}(k_1)
 i\Delta_{\phi^+}(-k_2)
 \left(ie m_W g_{\alpha'\mu}\right)\,
 v(p_+){\cal A}^\mu(q),
 \nonumber \\
 {\rm Im}\,\,tT_i(t)&=& \frac{g\sin^2\theta_W}{16\pi}
 \left[A_R^{(N)}-A_L^{(N)}\right]\frac{m_Wm_Z\sqrt{a_W}}{t-m_Z^2}.
 \end{eqnarray*}

\begin{flushleft}{\bf $\phi\phi Z$ polarization diagram}
\end{flushleft}
 \begin{eqnarray*}
 {\cal M}&=&\int d\varrho\,\,
 \overline{u}(p_-)\,
 \left[i(\Gamma^{(N)}_\nu+\Gamma^{(NC)}_\nu)\right]\,
 i\Delta_Z^{\nu\rho}(q) 
 \nonumber \\
 &&\times
 \left[-ig\frac{1-2\sin^2\theta_W}{2\cos\theta_W} 
                 (k_1-k_2)_\rho\right]
 \,i\Delta_{\phi^+}(-k_2)
 i\Delta_{\phi^-}(k_1)
 \Bigl[ie (k_1-k_2)_\mu\Bigr]\,
 v(p_+){\cal A}^\mu(q),
 \nonumber \\
 {\rm Im}\,\,tT_i(t)&=& \frac{g(1-2\sin^2\theta_W)}
                             {96\pi\cos\theta_W}
 \left[A_R^{(N)}-A_L^{(N)}\right]  \frac{a_W^{3/2}}{1-m_Z^2/t}.
 \end{eqnarray*}


\newpage
 \begin{table}\caption{C-, P-, T-properties of the spin, the 
                       electromagnetic field, and their interactions.}
              \label{CPT}
 \begin{center}
 \begin{tabular}{rccc} 
 
                                                 &  C  &  P  &  T  \\ 
                                                               \hline
 $\mbox{\boldmath $\sigma$}$                     & $+$ & $+$ & $-$ \\
 {\bf B}                                         & $-$ & $+$ & $-$ \\
 {\bf E}                                         & $-$ & $-$ & $+$ \\
 {\rm curl}\,{\bf B}, $\dot{{\bf E}}$            & $-$ & $-$ & $-$ \\
 $\mbox{\boldmath $\sigma$}\cdot{\bf B}$         & $-$ & $+$ & $+$ \\
 $\mbox{\boldmath $\sigma$}\cdot{\bf E}$         & $-$ & $-$ & $-$ \\
$\mbox{\boldmath$\sigma$}\cdot{\rm curl}\,{\bf B}$&$-$ & $-$ & $+$ \\
 $\mbox{\boldmath $\sigma$}\cdot\dot{{\bf E}}$   & $-$ & $-$ & $+$
 \end{tabular}
 \end{center}
 \end{table}
%
\newpage
\begin{figure}\mbox{\epsfig{file=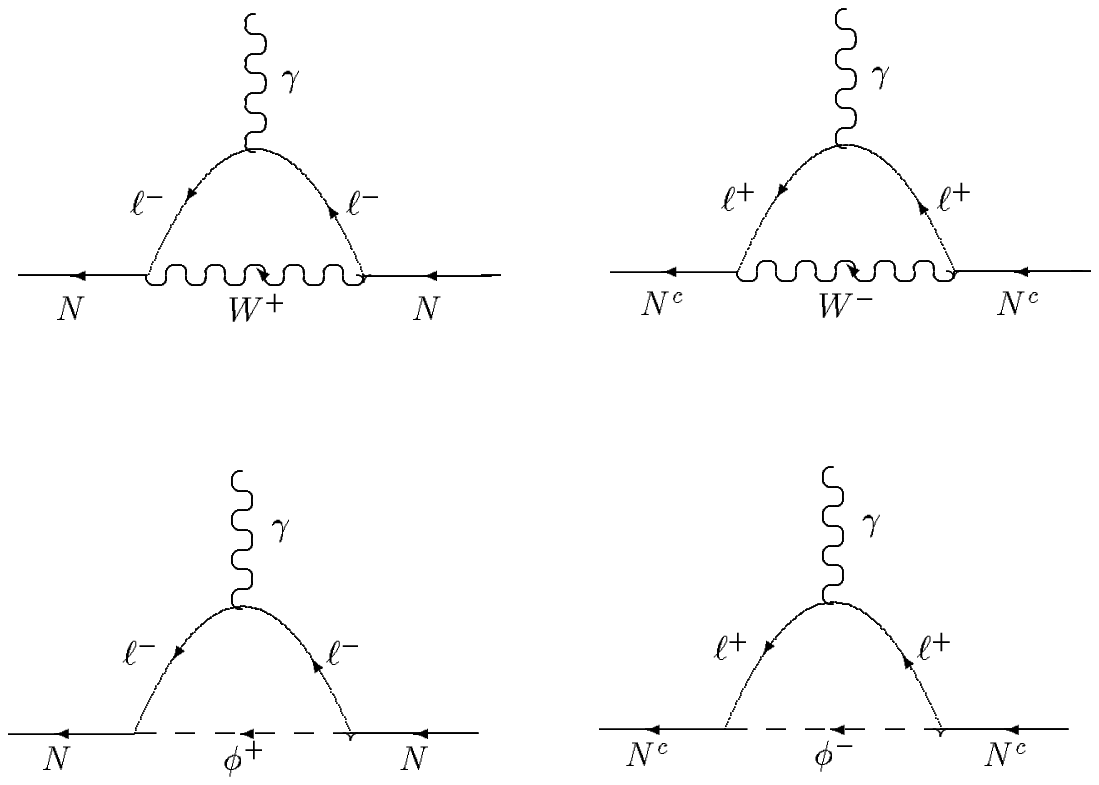,width=8.3cm}}\end{figure}
\begin{figure}\mbox{\epsfig{file=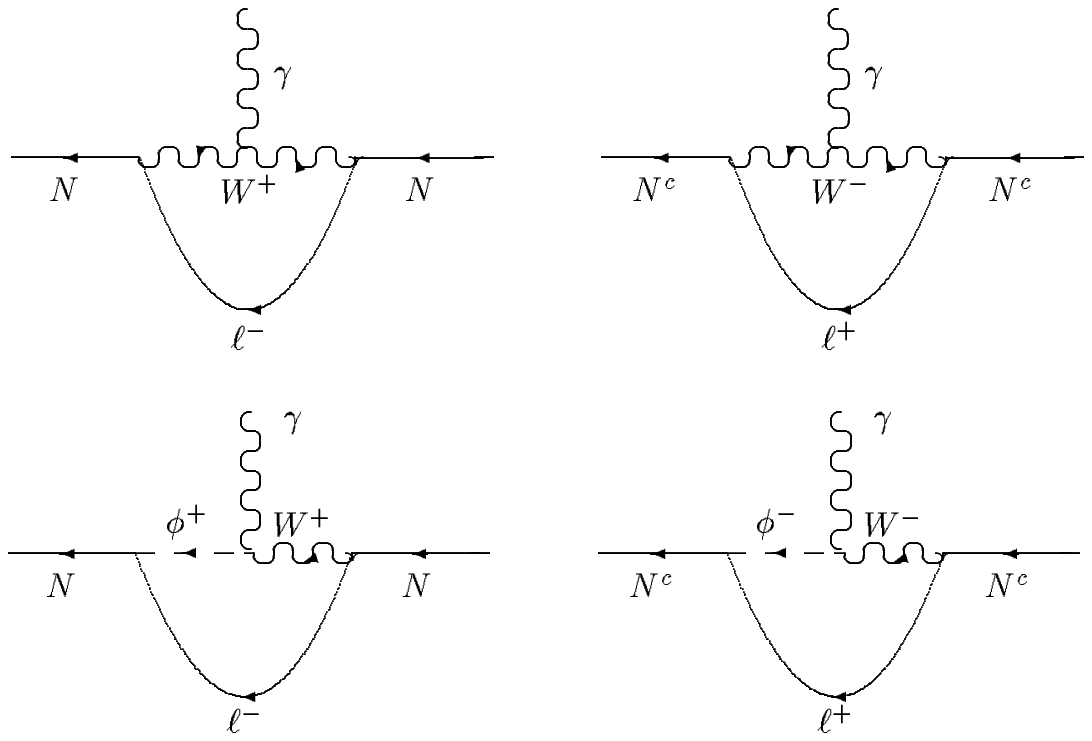,width=8.3cm}}\end{figure}
\begin{figure}\mbox{\epsfig{file=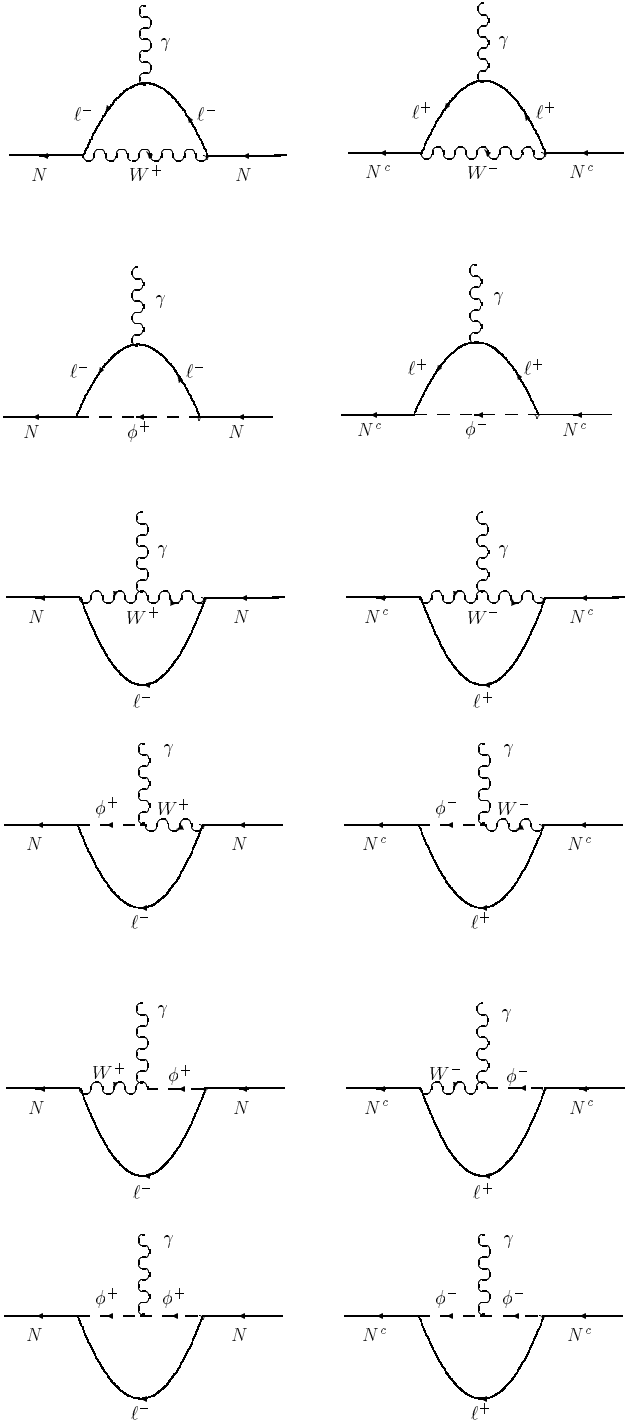,width=8.3cm}}
\vspace{0.5cm}
\protect\caption{\label{Fig1}
        Triangle diagrams responsible for the toroid moment of 
        the Majorana neutrino.}
\end{figure}
\newpage
\begin{figure}\mbox{\epsfig{file=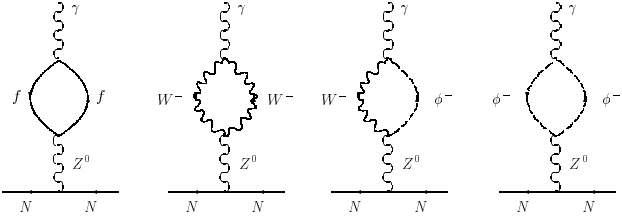,width=8.3cm}}
\protect\caption{\label{Fig2}
        Polarization-type diagrams responsible for the
        toroid moment of the Majorana neutrino.}
\end{figure}
\newpage
\begin{figure}\mbox{\epsfig{file=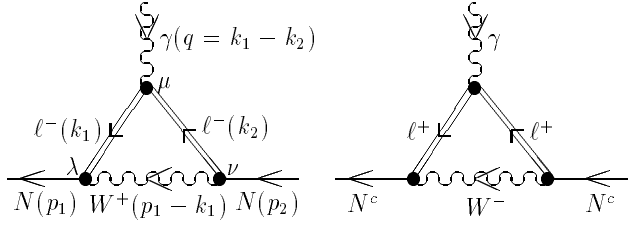,width=8.3cm}}
\vspace{-16cm}
\protect\caption{\label{Fig3}
        Feynman graphs with $\ell\ell W$ intermediate states.}
\end{figure}


\begin{references}
\bibitem{footnote2} For resonance conversion of neutrinos in the 
          medium, the MSW effect, the role of the matrix elements 
          ${\cal T}_{ii}$ plays the combinations 
          $(n_{\nu_\beta}-n_{\nu_\gamma})p_\nu$, where 
          $n_{\nu_\ell}$ is the neutrino index of refraction and 
          $p_\nu$ is the neutrino momentum.
\bibitem{footnote3} In all our calculations, we did not take into 
          account the widths of W,Z-bosons.
\end{references}
\end{document}